\documentclass[a4paper]{jpconf}
\usepackage{graphicx}
\begin{document}
\title{Overview and Status of the CMS Silicon Strip Tracker}

\author{Asish Satpathy}

\address{(\it {On Behalf of the CMS Tracker Collaboration}) \\ 
Department of Physics and Astronomy, University of California, Riverside, CA 92521, USA}
\ead{asish.satpathy@ucr.edu}

\begin{abstract}The CMS experiment at the LHC features the largest Silicon Strip Detector ever built. The impact of the operating conditions and physics requirements on the design choices of the CMS Silicon Tracker is reviewed. The readiness of the Silicon Strip Tracker for the tentatively scheduled CMS commissioning in Summer 2008 is briefly described.
\end{abstract}

\section{Introduction}
The Large Hadron Collider(LHC) at CERN is a 14 TeV proton-proton machine steadily getting ready to take data with a design luminosity of $10^{34}$ cm$^{-2}$ sec$^{-1}$, currently scheduled to start in Summer 2008. With an enormous potential physics reach, the Compact Muon Solenoid (CMS) \cite{cmsdet} detector is one of the two general purpose detectors at LHC that will independently detect some of Nature's the most interesting phenomena that we are yet to discover.


\section{Design Overview of the CMS Silicon Strip Tracker}
The LHC physics program requires an efficient and precise reconstruction of the tracks of the charged particles with a transverse momentum above 1 GeV in pseudo-rapidity ($|\eta|$) $<$ 2.5. Physics involving heavy flavors requires a precise measurement of secondary vertices and impact parameter for efficient particle identification. Efficient Lepton identification is required together with the CMS electromagnetic calorimeter and muon system. At the design luminosity, the tracker system is expected to operate efficiently in an environment where on average 1000 particles from more than 20 overlapping proton-proton interactions traversing the tracker every 25 ns. Being the closest detector to the interaction point, the system is expected to withstand an unprecedented level of radiation for an expected life of 10 years. All these stringent conditions place the tracker design concept into a challenging technological forefront. 


The CMS tracker \cite{tracker} is composed of a state-of-the-art pixel detector mounted close to the beam pipe and a silicon strip tracker outside the pixel detector. A total of 1,440 pixels modules are mounted in three barrel layers at radii between 4.4 cm and 10.2 cm and 2 end-cap disks on each side of the barrel. A total of 15,148 silicon strip modules are arranged in 10 barrel detection layers extending outward to a radius of 1.1 m and 9 disks on each side of the barrel, extending the acceptance of the tracker up to $|\eta| <2.5$. With about 200 m$^2$ active silicon area, the CMS tracker is the largest silicon tracker ever built.

The basic building block of the Silicon Strip Tracker is called a module. The tracker has 20 different module geometries at its various physical locations designed to operate at -10$^\circ$C. A module consists of four main parts: a support frame, a Kapton strip that delivers the bias voltage to the sensor back plane and insulates the sensor from support frame, the hybrids with front-end electronics and one or two sensors. The sensors are single-sided micro-strip detectors with AC-coupled p-type strips in an n-type bulk. Both single and double-sided modules are used. A double-sided module consists of two single-sided modules glued back to back with a stereo angle of 100 mrad.

Signals from the 512 or 768 silicon strips from a module are processed by four or six APV25 readout chips which are mounted on the front-end hybrid. APV25 \cite{APV} is a radiation hard 128 channels chip. Each of these channels consists of a preamplifier coupled to a shaping amplifier that produces a 50 ns CR-RC pulse shape. An analogue circuit for which two operation modes can be chosen processes signals from the level-one trigger. In 'peak' mode, only one data sample is used, while in 'deconvolution' mode three consecutive data samples are re-weighted and summed which gives much shorter pulse useful for correctly identify the bunch crossing during a high luminosity run of LHC. The signals of two APV25 chips are multiplexed onto one data line. The electrical signals are then converted to optical signals in dedicated Analogue-Opto Hybrids, and then digitized and processed by 10 bit ADCs.

The Silicon Strip Tracker is composed of three sub-detectors: Tracker Inner Barrel (TIB)/Tracker Inner Disk (TID), Tracker Outer Barrel (TOB) and Tracker End Cap (TEC). The TIB consists of four concentric cylinders placed at radii of 255.0 mm, 339.0 mm, 418.5 mm, and 498.0 mm, respectively, and extending from -700 mm to +700 mm along the z-axis. The two innermost layers have double-sided modules while the outer two layers have single-sided modules. The TID is composed of three disks per side, with three rings of modules per disk. TOB consists of a single mechanical cylindrical structure supporting 688 module sub-assemblies (from a total of 5200 Silicon modules) called rods. Each rod provides support and cooling for 6 or 12 silicon detector modules, together with their interconnection and readout electronics. The TOB has 6 layers of rods with average radii of 608, 692, 780, 868, 960, and 1080 mm. The cylinder has a total length of 2360 mm and inner and outer radii of 555 mm and 1160 mm respectively so that it goes right around the TIB/TID sub-detector. The TEC$+$ and TEC$-$ (two end caps) consist of wedge shaped carbon fiber support plate called petals which carries up to 28 modules arranged in 7 radial rings on each side. In addition to these seven disks, each of the end-caps contains 2 more disks serving as front/back termination. A total of 6400 modules are mounted on 288 petals to form 18 such disks for TEC$+$ and TEC$-$. The end-caps extend radially from 220 mm to 1135 mm and from $\pm$ 1240 mm to $\pm$2800 mm along the $z$-direction.

\section{Readiness of the Silicon Strip Tracker for Commissioning with Cosmics}
A semi-industrialized organization was developed for module production at various locations in the US and Europe in order to guarantee the uniform quality of the produced modules. Apart from regular diagnostic test, at each production site the modules were subject to long-term test (under bias and thermally cycled: -20$^{\circ}$C $<$ T $<$ +25$^{\circ}$C for 72 hours) to check their stability in time to qualify them for final integration. Module production for all three detectors components was completed by mid 2006 and then integrated onto petal and rod structures (in the case of TIB, directly onto half-shells). Fully tested rods and petals were then shipped to the integration center for final assembly to the support structure. TOB and TEC$+$ were integrated at a dedicated Tracker Integration Facility (TIF) at CERN. TIB/TID, TEC$+$ and TEC$-$ were integrated separately and later installed to the tracker support structure. The performance of every single module was compared and validated with data collected before and after integration. The layer integration of TIB/TID was completed in October 2006 and installation into the support tube was completed in January 2007. The performance of the TIB/TID was found to be excellent. In peak operation mode, signal to noise (S/N) ratio is measured to be about 28. The observed total number of dead channels is about 0.2\%. Figure 1 shows an example of noise distribution of the entire TOB after rod installation was completed in November 2006. Only less than a tenth of a percent of the total number of silicon strips are found either noisy or dead. The TOB modules have excellent S/N ratio that lie between 36-38 when operated in peak mode. Module production for TEC was completed in August 2006 and petal integration was completed in November 2006. TEC$+$ and TEC$-$ were installed to the tracker support tube (TST) in February 2007 and March 2007 respectively. This operation completes the assembly of the CMS tracker into the TST. The observed performance after TEC module integration was excellent with less than tenth of a percent of noisy strips and two tenth of a percent of dead channels.

\begin{figure}[h]
\begin{center}
\begin{minipage}{18pc}
\includegraphics[width=18pc]{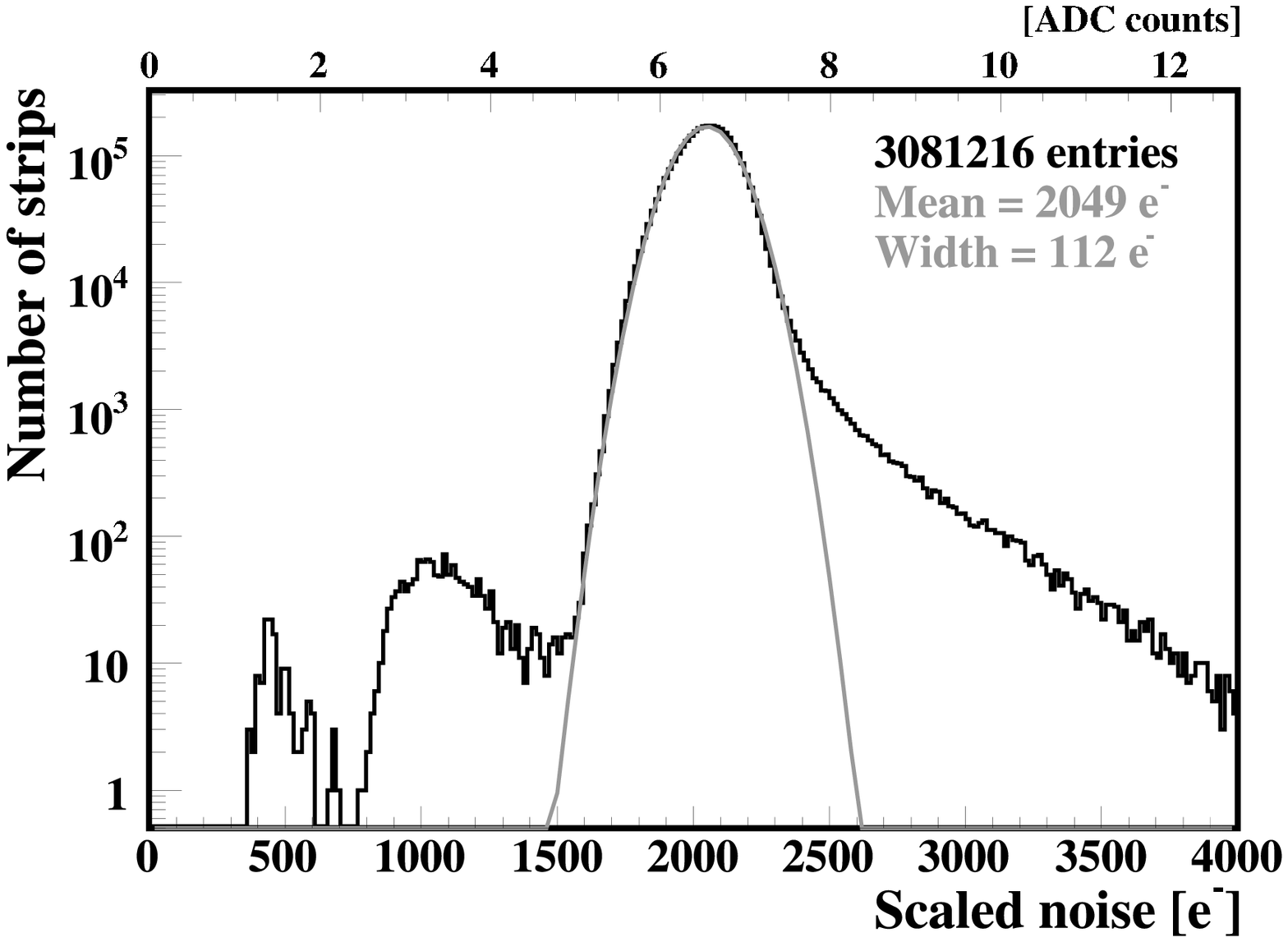}
\caption{\label{Noise}Common Mode Noise (in number of electrons) for all strips in the TOB after installation to the tracker support tube.}
\end{minipage}\hspace{2pc}%
\begin{minipage}{17pc}
\includegraphics[width=17pc]{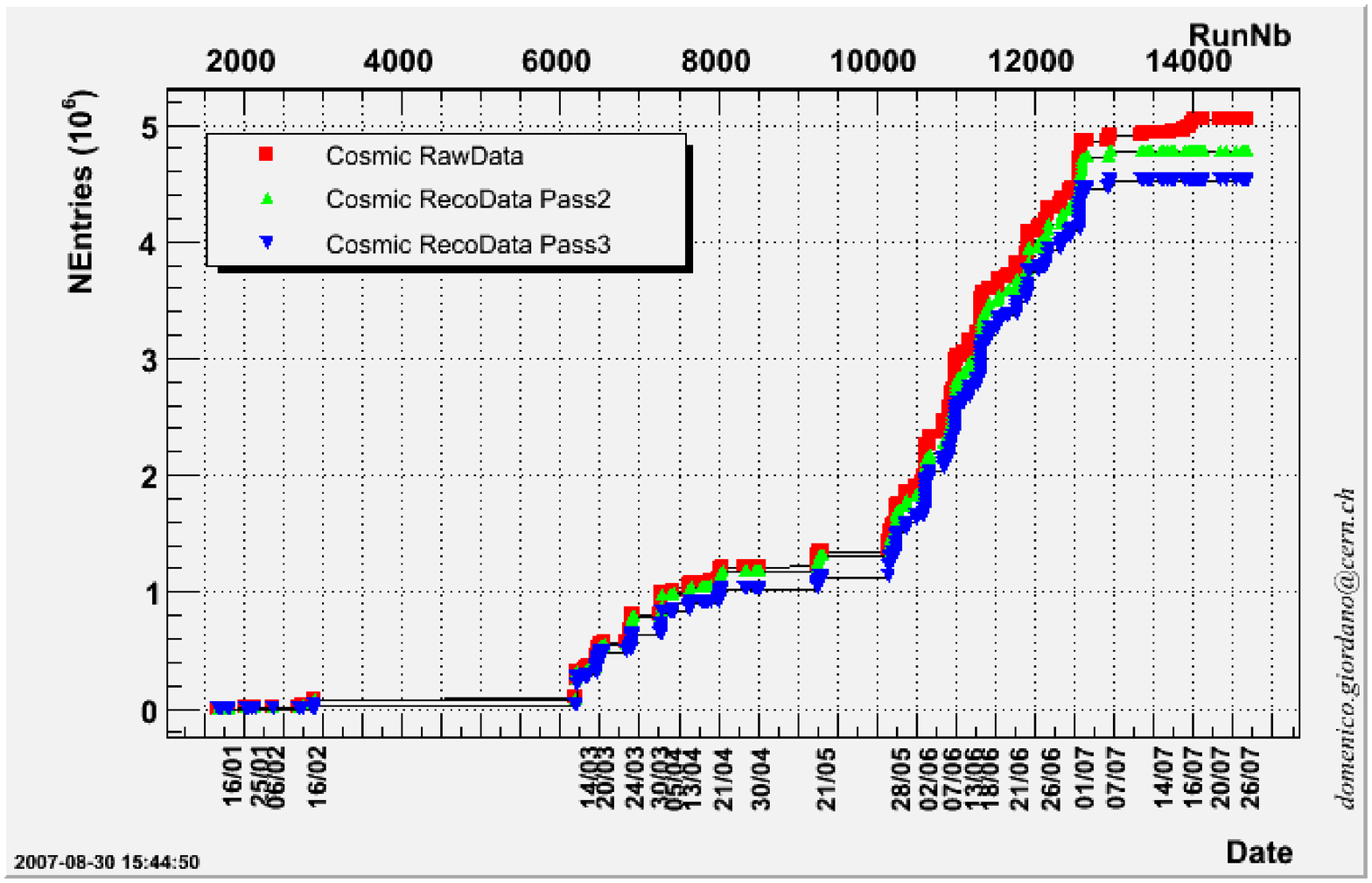}
\caption{\label{TIFData}Integrated Cosmic Triggered data by 25\% of the CMS Tracker at the TIF, CERN.}
\end{minipage} 
\end{center}
\end{figure}

The assembled CMS tracker as a single scientific instrument needs operational validation for which a slice of about 15\% of the detector was commissioned with cosmic trigger in an environment with close to an actual framework of the CMS software. As shown in Figure 2, nearly 5 million cosmic events were logged at various operating temperatures over a four months period. The cosmic trigger was chosen in such a way that different geometrical acceptance with the combination of TIB, TOB and TEC could be explored. A preliminary analysis of the slice test data has confirmed the expected goals of track parameters. Prior to the slice test, one percent of grade B tracker modules (4 rods, 2 TIB segments and 2 TEC petals) were tested in the CMS 4 T magnetic field with cosmic ray muons, and a lot of experience related to the detector hardware and operation (i.e. tracker service installation, geometry, slow control, safety, monitoring, tracking and software tools) was gained.

\section{Conclusion and Prospects}
Integration of the CMS silicon strip tracker in the support tube is completed and a very successful commissioning of 15\% of the tracker with cosmic triggers has just finished. In parallel with the preparatory work to move the tracker to the interaction point at the Point 5 cavern, a detailed systematic study of the data is underway. The overall performance of the tracker is excellent. Efforts are being made to deliver the tracker to the CMS with a goal of full commissioning with Cosmics before LHC turns on.

\end{document}